\documentclass[12pt]{article}

\usepackage{latexsym}

\font\tenmsbm=msbm10 scaled 1200
\font\sevenmsbm=msbm9
\newfam\msbmfam
\textfont\msbmfam=\tenmsbm
\scriptfont\msbmfam=\sevenmsbm

\makeatletter
\@addtoreset{equation}{section}
\makeatother

\newcommand{\eref}[1]{(\ref{#1})}

\def\beq{\begin{equation}}
\def\eeq{\end{equation}}
\def\bea{\begin{eqnarray}}
\def\eea{\end{eqnarray}}
\def\bet{\begin{tabular}}
\def\eet{\end{tabular}}
\def\pa{\partial}

\def\ve{\varepsilon}

\catcode`@=11

\begin{document}

\begin{titlepage}

\begin{flushright}
Preprint DFPD 01/TH/26\\
July 2001\\
\end{flushright}

\vspace{0.5truecm}

\begin{center}

{\Large \bf Anomaly free effective action for the elementary
$M5$--brane}

\vspace{2.5cm}

K. Lechner\footnote{kurt.lechner@pd.infn.it}, P.A.
Marchetti\footnote{pieralberto.marchetti@pd.infn.it} and M.
Tonin\footnote{mario.tonin@pd.infn.it}
 \vspace{2cm}

 {
\it Dipartimento di Fisica, Universit\`a degli Studi di Padova,

\smallskip

and

\smallskip

Istituto Nazionale di Fisica Nucleare, Sezione di Padova,

Via F. Marzolo, 8, 35131 Padova, Italia
}

\vspace{2.5cm}

\begin{abstract}
We construct an effective action describing an elementary
$M5$--brane interacting with dynamical eleven--dimensional
supergravity, which is free from gravitational anomalies. The
current associated to the  elementary brane is taken as a
distribution valued $\delta$--function on the support of the
5--brane itself. Crucial ingredients of the construction are the
consistent inclusion of the dynamics of the chiral two--form on
the 5--brane, and the use of an invariant Chern--kernel allowing
to introduce a $D=11$ three--form potential which is well defined
on the worldvolume of the 5--brane.

\vspace{0.5cm}

\end{abstract}

\end{center}
\vskip 2.0truecm
\noindent PACS: 11.15.-q, 11.10.Kk, 11.30.Cp;
Keywords: M5--branes, anomalies, singular currents.
\end{titlepage}

\newpage

\baselineskip 6 mm


\section{Introduction}

The until now only conjectured $M$--theory is supposed to be a
unifying consistent theory in eleven dimensions whose low energy
limit is $D=11$ supergravity. Its elementary excitations are
2--branes and 5--branes which are ``electromagnetically" dual to each
other. These two excitations can coexist if their charges $e$ and
$g$ satisfy the Dirac--quantization condition
 \beq
 \label{Dirac}
 eg=2\pi n\, G,
 \eeq
where $G$ is the eleven--dimensional Newton's constant,
usually written as $G=2\kappa^2$, and $n$ is an integer.

The dynamics of the bosonic sector of the $M2$--brane is described
by the coordinates $x^\mu(\sigma)$, $\mu=0,\cdots,10$, and the
worldvolume swept out during its time evolution is
three--dimensional. The bosonic sector of an $M5$--brane is
described by the coordinates $x^\mu(\sigma)$ and by the
self--interacting chiral two--form $b_{ij}(\sigma)$, whereas its
worldvolume is six--dimensional. Thus, the main differences
between the two excitations are the presence of the two--form
$b_2$ and the possibility of gravitational anomalies on the
5--brane, while 2--branes are trivially anomaly free.

As shown in \cite{Witten2} the gravitational anomaly generated by
$b_2$ and by the two complex chiral fermions on the 5--brane is
represented by the anomaly polynomial $2\pi(X_8^{(0)}+{1\over 24}P_8)$,
with
 \bea
 X_8&=&{1\over
192\,(2\pi)^4}\left(tr\,R^4-{1\over4}(tr\,R^2)^2\right)
\label{target}\\
P_8&=&{1\over 8\, (2\pi)^4}\left((tr\,F^2)^2-2\,
tr\,F^4\right), \label{Pont}
 \eea
where $R$ is the target space
$SO(1,10)$--curvature and $F$ the $SO(5)$--curvature of the normal
bundle of the 5--brane. With $X_8^{(0)}$ we denote the pullback 
of the target space polynomial $X_8$ on the $M5$--brane worldvolume.
The target space anomaly, associated to $X_8$, can be cancelled \`a la
Green--Schwarz modifying the equation of motion of the $D=11$
four--form curvature $H_4$, while $P_8$, the second Pontrjagin
form, represents the residual anomaly whose cancellation requires
(some sort of) the inflow mechanism. The anomaly itself, as
variation of the quantum effective action $\Gamma_q$, is obtained
through the descent formalism,
 \beq
 \label{grav}
 {\cal A}=\delta\, \Gamma_q=2\pi\int_{M_6}\left(X_6^{(0)}+
   {1\over 24}P_6\right),
 \eeq
where $M_6$ is the 5--brane worldvolume. Our notation for descent
equations is $X_8=dX_7, \,\delta X_7=dX_6$, and similarly for
$P_8$. $X_6^{(0)}$ denotes again the pullback of $X_6$ on $M_6$.

The fundamental equation which describes the coupling of a
5--brane with charge $g$ to eleven--dimensional supergravity is
 \beq
 \label{j5}
dH_4=gJ_5,
 \eeq
where the 5--form $J_5$ is essentially the Dirac
$\delta$--function on the 5--brane worldvolume (see below for a
precise definition); we refer to such branes  carrying a
current with $\delta$--like support as {\it elementary} branes.
It is eventually this equation which should induce the
cancellation of the residual $SO(5)$--anomaly through inflow. In
pure supergravity one has $dH_4=0$, and this allows to introduce a
potential through $H_4=dB_3$. If on the other hand $g\neq 0$, the
first problem one has to face is how to introduce  a potential
$B_3$ in a consistent way. Since, moreover, the action for pure
supergravity is cubic in $B_3$, the presence of a current $J_5$
with $\delta$--like support leads in the action to cubic products
of terms with at least inverse--power--like short distance singularities;
the second problem one has to face is related with an accurate treatment
of these singularities.

There have been various attempts to deal consistently with
equation \eref{j5}, with the aim of cancelling the residual
gravitational  anomaly. To circumvent the second problem, the
strategy adopted in Ref. \cite{Harvey} consists in smoothing out
the singular source $J_5$ and to replace it with a specific
regular one, $J_5^{reg}$, carrying the same total flux as $J_5$.
With this choice for the current the authors of \cite{Harvey} were
able to construct a modified Wess--Zumino term, replacing
${1\over 6}\int B_3dB_3dB_3$ of pure supergravity, whose
variation cancels indeed the residual anomaly. A drawback 
of a regular current $J_5^{reg}$ is that it does not admit
a consistent coupling to elementary $M2$--branes: since
the 5--brane charge is now continuously  distributed Dirac's
condition \eref{Dirac} is no longer sufficient to make the
Dirac--brane associated to the $M2$--brane unobservable.
A Dirac--brane associated to the
$M2$--brane is a 3--brane whose boundary is the $M2$--brane; it
represents a generalization of the Dirac--string of a
four--dimensional monopole. If $M2$--branes and $M5$--branes are
simultaneously present the introduction of at least one
Dirac--brane is unavoidable, in complete analogy with the case of
charges and monopoles in four dimensions, see e.g. ref.
\cite{LM}. A part
from this one should explain why the regular current associated to
the 5--brane should have the particular form $J_5^{reg}$. The
authors of \cite{Bonora} instead insist on a $\delta$--like
current and argue, as a consequence of eq. \eref{j5}, that the
5--brane $SO(5)$--normal bundle $N$ splits in a line bundle $L$
and an $SO(4)$--bundle $N^\prime$. This allows them to consider in
the residual anomaly polynomial only $SO(5)$--connections which
are reducible to $SO(4)$--connections, and to construct a local
counterterm which cancels the corresponding anomaly. However,
there remains an unphysical dependence on the choice of the splitting.
Notice also that both references do not worry about the dynamics
of the $b_2$--field. Finally, the cancellation of the residual anomaly
in the compactified theory, corresponding to an $NS5$--brane in
$D=10$, $IIA$--supergravity, has been realized in
\cite{Witten2,Becker}.

Aim of this paper is the construction of the low energy
dynamics of the bosonic elementary $M$--theory 5--brane, coupled
to the bosonic sector of dynamical $D=11$ supergravity; the
cancellation of the residual anomaly will be an automatic output
of our construction, rather than an a priori requirement. Our
point of view is that if $M$--theory is a consistent theory, there
should exist a consistent low energy dynamics describing the
interaction of $M5$-- and $M2$--branes with/through {\it
dynamical} eleven--dimensional supergravity. In this sense our
approach goes beyond the $\sigma$--model approach where the target
space fields are supposed to satisfy the equations of pure
source--less supergravity. We will concentrate on the dynamics of
the 5--brane, since it bears the major difficulties, and include
the 2--brane only at the end. Crucial ingredients of the
construction are the inclusion of the $b_2$--field dynamics, and a
consistent solution of \eref{j5} in terms of a $D=11$ three--form
potential which admits a well defined pullback on $M_6$, i.e.
which is regular in the vicinity of the 5--brane worldvolume.
There is a standard approach \cite{LM} to solve such an
equation, involving Dirac--branes. In the present case however, due to the
cubic interactions in the action, we need an alternative 
approach in terms
of Chern--kernels \cite{Chern,charact}, which are able to codify
the physical singularities of $H_4$ near the 5--brane in a universal way.

Since we insist on a $\delta$--like current our natural framework
is the one of $p$--currents (rather than $p$--forms), i.e. of
$p$--forms with distribution valued coefficients \cite{deRham};
consequently the
differential $d$ acts always in the sense of distributions,
otherwise an equation like \eref{j5} would never make sense. We
suppose also that our eleven--dimensional target space $M_{11}$ is
topologically trivial, so every closed $p$--current is also exact.
Henceforth we will call our ``currents" again ``forms".

The present paper presents the main result, i.e. the anomaly free
low energy effective action, eqs. \eref{action}, \eref{kin}
and \eref{L11}; detailed proofs and applications will be presented 
elsewhere \cite{fullpaper}.

\section{Equations of motion}

The bosonic fields of $D=11$ supergravity are the metric
$g_{\mu\nu}(x)$ and the three--form potential $B_3$; the bosonic
fields on the closed 5--brane are the coordinates
$x^{\mu}(\sigma)$, $\sigma^i=(\sigma^0,\cdots,\sigma^5)$ and the
two--form $b_2$. The field $B_3$ can also be dualized to a
six--form $B_6$, but since there exists no formulation of $D=11$
supergravity which involves only $B_6$ it is preferable to use a
formulation in terms of only $B_3$. We indicate the curvatures
associated to $B_3$ and $b_2$ respectively as $H_4=dB_3+\cdots$
and $h_3=db_2+\cdots$. With the upper index $(0)$ we will indicate
the pullback of a target space form to the 5--brane worldvolume
$M_6$ whenever it exists, e.g. $B_3^{(0)}$ indicates the pullback
of $B_3$ to the six--dimensional submanifold $M_6$.

We propose, as starting point, the following set of classical
equations of motion and Bianchi identities  for $H_4$ and $h_3$,
 \bea
h_3&=&db_2+B_3^{(0)} \label{c}\\
h_{ij}&=&-2\,{\delta {\cal L}\over \delta \widetilde h_{ij}}\label{d}\\
dH_4&=&gJ_5\label{a}\\
d*H_4&=&{1\over 2}\,H_4H_4+g\,h_3\circ J_5+{2\pi G\over
g}\,X_8\label{b}.
 \eea
In equation \eref{c} we defined the curvature of the two--form
potential $b_2$, according to the $\sigma$--model approach, in
terms of the pullback of the target space potential $B_3$. This
implies first of all that this pullback has to be well defined.
Equation \eref{d} amounts then to the generalized self--duality
equation of motion for $b_2$ which is induced by the Born--Infeld
lagrangian ${\cal L}(\widetilde h)= \sqrt
{det(\delta_i{}^j+i\,\widetilde h_i{}^j)}$ where, according to the
PST--approach \cite{PRL,dbipst}, $h_{ij}=v^kh_{ijk}$, $\widetilde
h_{ij}=v^k(*h)_{ijk}$, $v_k=\pa_ka/\sqrt{-(\pa a)^2}$, and
$a(\sigma)$ is a non propagating scalar auxiliary field.

Equation \eref{a} is a Bianchi identity for $H_4$ which has to be
solved  in terms of the potential  $B_3$; after that \eref{b}
becomes an equation of motion for this field. The five--form $J_5$
in \eref{a} is defined as the Poincar\'e dual in the space of
currents \cite{deRham} of the five--brane worldvolume, i.e. $\int
\Phi_6J_5 =\int_{M_6} \Phi_6^{(0)}$ for every smooth target space
six--form $\Phi_6$. In an arbitrary coordinate system a local
expression for $J_5$ is
 \beq
 \label{expl}
 J_5={1\over 5!6!}\,
dx^{\mu_1}\cdots
dx^{\mu_5}\,\ve_{\mu_1\cdots\mu_5\nu_1\cdots\nu_6} \int_{M_6}
E^{\nu_1}\cdots E^{\nu_6}\,\delta^{11}(x-x(\sigma)), \label{basic}
 \eeq
 where $E^\mu(\sigma)=d\sigma^i E^\mu_i(\sigma)$, and the
$6$ vectors $E^\mu_i(\sigma)=\pa_ix^\mu(\sigma)$ form a basis for
the tangent space on $M_6$ at $\sigma$.

A basic problem one has to solve is how to give a well--defined
meaning to the r.h.s. of \eref{b}, as a {\it closed target space
eight--form}. For $g=0$ it reduces to the $B_3$--equation of
motion of pure supergravity \footnote{As we will see below, the
5--brane charge is related to Newton's constant by $2\pi G=g^3$.}.
The second and third term on its r.h.s. are dictated by the
presence of the 5--brane; the term proportional to $X_8$, see
\cite{X8}, realizes the standard Green--Schwarz cancellation
mechanism for the target space anomaly and is a closed form. The
presence of the second term -- indications for its appearance have
been given first in \cite{h3j5} -- is needed to make the r.h.s. of
\eref{b} a closed form, see below. In the form \eref{b} this
formula has been proposed in \cite{Witten3}.

We begin by specifying what we mean with the expression $h_3\circ
J_5$: technically this target space eight--form is defined as the
canonical push--forward of the 5--brane field $h_3$ to the
eleven--dimensional target space. In general the ordinary product
between a form on the brane and a form on the target space defines
neither a form on the brane, nor a form  on the target space.
However, a "product" of the kind $h_n\circ J_p$, where $J_p$ is
the Poincar\'e dual of a $(D-p)$--manifold and $h_n$ an $n$--form
on that manifold $(n+p\leq D)$, is defined in the distributional sense as a
target space $(n+p)$--form  according to
 \begin{equation}
\int_{R^D}\Phi_{D-n-p}\,(h_n\circ J_p)=\int_{M_{D-p}}
\Phi_{D-n-p}^{(0)}\,h_n,
 \end{equation}
for every test form. A local expression, following from this
definition, is
 \bea
 h_n\circ J_p &=& {1\over (n+p)!(D-n-p)!}\,
 dx^{\mu_1} \cdots dx^{\mu_{n+p}}
\,\varepsilon_{\mu_1\cdots \mu_{n+p}\nu_1\cdots\nu_{D-n-p}}\\
& &\cdot \int_{M_{D-p}} E^{\nu_1}\cdots
E^{\nu_{D-n-p}}\,h_n\,\delta^D(x-x(\sigma)).
 \eea
This corresponds to a local expression for the push--forward of
$h_n$, and it involves only the worldvolume field $h_n$ and none
of its "by hand" extensions. The product notation $h_n\circ J_p$
is useful because from the above definition it follows that the
standard Leibnitz rule holds for push--forward forms "as if they
were factorized",
$$
d(h_n \circ J_p)=h_n\circ dJ_p+(-)^p dh_n\circ J_p.
$$
The last property of the push--forward operation we need is
$$
\Phi\, J_p= \Phi^{(0)}\circ J_p\, ,
$$
for every target space form $\Phi$ which admits pull back.

The next point concerns the term ${1\over 2}H_4H_4$ at the r.h.s.
of \eref{b}, and its differential. The problematic aspect of this
eight--form is represented by the fact that, due to \eref{a},
$H_4$ exhibits necessarily singularities near $M_6$, meaning that
$H_4^{(0)}$ does not exist; in particular the computation
$d({1\over 2}H_4H_4)=gH_4J_5=g H_4^{(0)}\circ J_5$ makes no sense.

To settle the question of how to compute the differential of
$H_4H_4$ one must first specify the singularities near $M_6$
present in $H_4$. Since $J_5$ is closed we can always write
$J_5=dK_4$ for some four--form $K_4$, and then $H_4=dB_3+gK_4$;
the singularities of $H_4$ are then the ones of $K_4$ because $B_3$ is 
regular. Since $J_5=dK_4$ these singularities can be essentially of 
two types: the first type corresponds to $\delta$--like singularities
induced by a Dirac--brane, i.e. a 6--brane whose
boundary is the 5--brane. This would lead to $K_4=C_4$, where
$C_4$ is the delta--function on the Dirac--brane, i.e. its
Poincar\'e--dual. But for such singularities the product $H_4H_4$
would not even define a distribution since the square of a 
$\delta$--function is not defined 
\footnote{Viceversa, if you 
require that the combination $H_4=dB_3+gC_4$ does not exhibit
$\delta$--like 
singularities then $B_3$ can not be regular near $M_6$ because
$dB_3$ must cancel the $\delta$--function singularities in $C_4$.}. 
For this reason the Dirac--brane approach can not be applied to the
$M5$--brane effective action, based on the system
\eref{c}--\eref{b}.

The second type of possible singularities is represented by
inverse--power--like singularities, like the ones of a
Coulomb--field whose divergence equals a $\delta$--function,
supported on the position of the source. In this case the
inverse--power--like singular behaviour of $K_4$ near the 5--brane
should be {\it universal}. By definition this  behaviour is realized by the
Chern--kernel \cite{Chern}, see below, which is appropriately expressed in
terms of normal coordinates. Since the Chern--kernel will lead
also to a well--defined product $H_4H_4$ and, eventually, to a
r.h.s. of \eref{b} which defines a closed target space
eight--form, we will base our effective action on this kernel.

\subsection{Normal coordinates and Chern--kernels}

We regard the introduction of a system of normal coordinates
as a $D=11$ diffeomorphism from the coordinates $x^\mu$ to
the coordinates $(\sigma^i,y^a)$, with $i=0,\cdots,5$ and $a=1,\cdots,5$,
specified by the functions $x^\mu(\sigma,y)$. The coordinates $y^a$ are
called ``normal" in that we require that
 \beq \label{normal}
x^\mu(\sigma,0)=x^\mu(\sigma),\quad
N_\mu^a E^\mu_i=0,\quad N_\mu^a N^{\mu b}=\delta^{ab},
 \eeq
where
$$
N^\mu_a(\sigma)\equiv \left.{\pa x^\mu(\sigma,y)\over \pa
y^a}\right|_{y=0}.
$$
As a power series in $y$ we have therefore
 \beq
 \label{series}
x^\mu(\sigma,y)=x^\mu(\sigma)+y^aN^{\mu a}(\sigma)+o(y^2).
 \eeq
Since the  vectors $N^\mu_a(\sigma)$ specify a basis for the
normal fiber, $SO(5)$--connection and curvature on $M_6$ can be
parametrized by
 \beq
 \label{connection}
A^{ab}=N^{\mu b}\left(dN_\mu^a+\Gamma_\mu^\nu N_\nu^a\right),\quad
F^{ab}=dA^{ab}+A^{ac}A^{cb},
 \eeq
where $\Gamma$ is the pullback of the eleven--dimensional affine
connection.

Notice that, for chosen $N^{\mu a}$, the conditions \eref{normal}
determine only the structure of the coordinate system near the
5--brane; away from the 5--brane the coordinate system is only
required to be one to one. So there is a large freedom left, which
is expressed by the $o(y^2)$--terms above. For simplicity we
suppose here that the normal coordinate system is defined globally
in target space; the adaptation of our construction to the general
case, where it can be defined only locally, is sketched in section
five.

The definition of a Chern--kernel with the correct fall--off at
infinity requires also the introduction of an extended
$SO(5)$--connection one--form $A^{ab}(\sigma,y)$ on the whole
target space, asymptotically flat in $|y|$ and restricted by the boundary
conditions
 \beq
 \label{ext} A^{ab}(\sigma,0)=A^{ab}(\sigma).
 \eeq
This means that the pullback of $A^{ab}(\sigma,y)$ on the 5--brane
reduces to the $SO(5)$--connection defined in \eref{connection},
and that its curvature goes to zero at infinity along all
$y$--directions. Unless otherwise stated from now on we will
always use this extended connection and the associated extended
curvature $F^{ab}$.

The systems of normal coordinates and of extended connections fall
into $SO(5)$--equivalence classes, the representatives being
related by local $SO(5)$--transformations
$\Lambda^{ab}(\sigma,y)$,
$$
\widetilde y^a=\Lambda^{ab}y^b,\quad \widetilde A=\Lambda
A\Lambda^T -\Lambda d \Lambda^T.
$$

In terms of an arbitrary normal coordinate system the current
$J_5$ admits the simple local expression
 \beq\label{j5normal}
J_5={1\over 5!}\,dy^{a_1}\cdots dy^{a_5}\,\ve^{a_1\cdots
a_5}\,\delta^5(y),
 \eeq
and we can now ask if there exists an $SO(5)$--invariant
four--form $K_4$, polynomial in $\hat y^a\equiv y^a/\sqrt{y^2}$
and $A^{ab}$, satisfying
 \beq\label{kernel}
J_5=dK_4.
 \eeq
Such a four--form exists, it is indeed uniquely determined, and it
is expressed in terms of the above data by the Chern--kernel
\cite{Chern,charact},
 \beq\label{k4}
K_4={1\over 16(2\pi)^2}\,\ve^{a_1\cdots a_5}\,\hat
y^{a_1}K^{a_2a_3}K^{a_4a_5},
 \eeq
where
 $$
 K^{ab}\equiv F^{ab}+D\hat y^a D\hat y^b,\quad
 D\hat y^a=d\hat y^a+\hat y^bA^{ba}.
 $$
Local $SO(5)$--invariance is manifest and to verify \eref{kernel}
one has to compute the differential of $K_4$ in the sense of
distributions \footnote{The unique non vanishing contribution in
the differential of $K_4$ comes entirely from\par\noindent
$d\left({1\over 16(2\pi)^2}\,\ve^{a_1\cdots a_5}\, \hat
y^{a_1}d\hat y^{a_2}d\hat y^{a_3}d\hat y^{a_4}d\hat
y^{a_5}\right)=J_5$.}. The salient properties of this four--form
are that far away from the 5--brane, $y^a\rightarrow \infty$, it
exhibits a typical Coulomb--like behaviour $K_4\sim {1\over
16(2\pi)^2}\,\ve^{a_1\cdots a_5}\, \hat y^{a_1}d\hat y^{a_2}d\hat
y^{a_3}d\hat y^{a_4}d\hat y^{a_5}$, while near the 5--brane,
$y^a\rightarrow 0$, it exhibits a universal $SO(5)$--invariant
behaviour, which is independent of the choice of normal
coordinates and of the extension of $A$. Notice, however, that the
pullback of $K_4$ on $M_6$ does not exist \footnote{The four--form
$K_4$ has been introduced, as $1/2\,e_4$, also in \cite{Harvey}
but there it was treated as a closed form as it is away from the 
5--brane.}.

We must stress that, although $K_4$ depends only on the equivalence
class of normal coordinate systems and extended
$SO(5)$--connections, it changes if one chooses {\it another}
equivalence class. Inequivalent systems of normal coordinates
are related by a transformation $y^a\rightarrow y^{\prime a}(\sigma,y)$,
such that
\beq
\label{struc}
 y^{\prime a}(\sigma,0)=0 \qquad {\partial y'^a \over \partial y^b}
(\sigma,y)|_{y=0} = \delta^{ab}.
\eeq
Such a change corresponds precisely to the ambiguity
associated to the $o(y^2)$--terms in \eref{series}, which, in
turn, reflect the huge arbitrariness of the normal coordinate
systems away from the 5--brane. Moreover, one can choose
infinitely many different extensions of the $SO(5)$--connection
$A(\sigma)$ from a form on $M_6$ to a target space form,
compatible with $y$--asymptotic flatness and \eref{ext}. Under both types
of changes we obtain
a different four--form $K_4^\prime$ such that
 $$
 dK_4^\prime=J_5=dK_4;
 $$
Poincar\'e's lemma implies then that locally there exists a
three--form $Q_3$ such that
 \beq
 \label{changeclass}
K_4^\prime=K_4+dQ_3.
 \eeq
Moreover, since $K_4^\prime$ and $K_4$ carry the {\it same}
singular behaviour near the 5--brane, $Q_3$ behaves regularly as
$y^a\rightarrow 0$ and using \eref{ext} and \eref{struc} one
can verify that it has vanishing pullback on $M_6$,
 \beq\label{q3}
 Q_3^{(0)}=0.
 \eeq
This piece of information will become important in a moment. Since
$K_4$ is $SO(5)$--invariant, we can now introduce an
$SO(5)$--invariant three--form potential $B_3$ according to
 \beq
 \label{b3}
H_4=dB_3+gK_4.
 \eeq
Under a change of equivalence class \eref{changeclass} we must
require
 \beq
\label{boh}
B_3^\prime=B_3-gQ_3,
 \eeq
such that $H_4$ is independent of the new structures
that we have introduced to construct $K_4$, i.e. the  particular
normal coordinate system that we have chosen and the particular
extension of the $SO(5)$--connection. Notice also that \eref{q3}
ensures that $B_3^{(0)}$ as well as $h_3$, apart from being
well--defined, are independent of the new structures, too.
 Equation \eref{b3} provides a splitting of $H_4$ into a
regular part which is also closed, $dB_3$, and a singular part,
$K_4$, with a universal behaviour near $M_6$, in view of
\eref{changeclass} and \eref{q3}.

The form $K_4$ satisfies the following chain of relations 
 \bea
 dK_4&=&J_5\label{first}\\
 d\left(K_4K_4\right)&=&0\label{second}\\
 d\left(K_4K_4K_4\right)&=&{1\over 4}\,P_8\,J_5\label{third}\\
 d\left(K_4K_4K_4K_4\right)&=&0\label{fourth},
 \eea
where $P_8$ is the second Pontrjagin form. These
relations follow from an identity whose proof we will present in
\cite{fullpaper} (see however also \cite{Intri} and
\cite{BottCatt}):
 \beq
 \label{fund}
 K_4K_4={1\over 4}\,df_7,\quad f_7=P_7+Y_7,
  \eeq
where $P_7$ is the Chern--Simons form associated to the Pontrjagin
form $dP_7=P_8$, and $Y_7$ is an $SO(5)$--invariant seven--form
given by
 $$
Y_7={1\over (2\pi)^4}\left[\hat y^aD\hat y ^b
(F^3)^{ba}+\left({1\over
 2}\,tr\,F^2-D\hat y^cD\hat y^dF^{cd}\right)\hat y^aD \hat
 y^bF^{ab}\right].
 $$
This proves immediately \eref{second}. To prove \eref{third} one
has also to use that in the sense of distributions
 \beq
 \label{id}
d\left(Y_7K_4\right)=dY_7K_4.
 \eeq
 Notice that, due to the singular behaviour of $K_4$ near the
5--brane, one is not allowed to use Leibnitz's rule for
differentiation; otherwise in the above formulae one would obtain
some meaningless expressions like $K_4J_5$ and $Y_7J_5$.

Formula \eref{fund} means that the inverse--power--like singularities
of $K_4$ which give rise to the $\delta$--function
in $dK_4$, cancel in the product $K_4K_4$ due to antisymmetry reasons,
and that $K_4K_4$ amounts to a closed eight--current.
Using this formula it is finally easy to verify that the
r.h.s. of \eref{b} is a well--defined closed form. It suffices to
notice that
 \bea
\nonumber d\left({1\over 2}H_4H_4\right)&=&{1\over
2}d\left(dB_3dB_3+2gdB_3K_4+g^2 K_4K_4\right)\\
                          &=& gdB_3J_5=g dB_3^{(0)}\circ J_5,
                          \label{conto}
 \eea
which cancels against $d(gh_3\circ J_5)=-gdh_3\circ J_5=-g
dB_3^{(0)}\circ J_5$.

Since we have now a well defined system of equations of motion we
can search for an action which gives rise to it. This is the aim of the
last section.

\section{The effective action}

We write the bosonic effective action $\Gamma$ for an $M5$--brane 
with charge $g$ interacting with $D=11$ supergravity 
as the sum of a local classical action, which should reproduce the
equations of motion for $b_2$ and $B_3$, resp. \eref{d} and
\eref{b}, and of the quantum effective action,
 \beq \label{action}
 \Gamma= {1\over G}\left(S_{kin}+S_{wz}\right)+\Gamma_q,
 \eeq
where we separated the classical action in kinetic terms and in a
Wess--Zumino action. The invariant curvatures are given in
\eref{c} and in \eref{b3}, so the reconstruction of the classical
action is, indeed, a merely technical point. Actually, the field
equations for $B_3$ and $b_2$ fix the classical action modulo
terms which are independent of these fields; these terms are, in
turn, fixed by invariance requirements, in the present case independence
of the action of the choice of normal coordinates and of the extension of
the $SO(5)$--connection. More precisely, according to the previous section
we have to require invariance under
 \bea K_4^\prime&=&K_4+dQ_3\label{1}\\
      B_3^\prime&=&B_3-gQ_3\label{2}\\
           f_7^\prime&=&f_7 +8K_4Q_3 +4Q_3dQ_3+dQ_6\label{2a}\\     
       Q_3^{(0)}&=&0=Q_6^{(0)}.\label{3}
 \eea
The relation \eref{2a} follows from the definition of $f_7$ in 
\eref{fund} and from the relation $K_4^\prime K_4^\prime=
{1\over4}df_7^\prime$. It determines the seven--form     
$f_7^\prime\equiv P_7^\prime +Y_7^\prime$ modulo a closed form 
$dQ_6$. The pullback of $Q_6$ vanishes for the same reasons as 
the pullback of $Q_3$. 

Clearly, in the absence
of the 5--brane we want to get back the action of pure $D=11$
supergravity. Employing for the two--form field equation \eref{d}
the covariant PST--approach \cite{dbipst}, the invariant kinetic
terms for the space--time metric, for $B_3$, $b_2$ and
$x^\mu(\sigma)$ are given by
 \beq
 \label{kin}
S_{kin}=\int_{M_{11}} d^{11}x\sqrt{g_{11}}\,R-{1\over
2}\int_{M_{11}}H_4*H_4 -g\int_{M_6}d^6\sigma\sqrt{g_6}\left({\cal
L}(\widetilde h)+{1\over 4} \widetilde h^{ij}h_{ij}\right),
 \eeq
where $g_6$ is the determinant of the induced metric on the
5--brane. Notice that $H_4$ as well as $h_3$ are manifestly invariant
under \eref{1}--\eref{3}.

The Wess--Zumino action, which appears to be
the crucial ingredient of the effective action, is written as the integral
of an eleven--form, $S_{wz}=\int_{M_{11}}L_{11}$, with
 \bea
\label{L11}
 L_{11}&=&{1\over 6}\,B_3dB_3dB_3 -{g\over 2}\left(b_2\,dB_3^{(0)}\right)
 \circ J_5+{g\over 2}\,B_3dB_3K_4 + \nonumber \\
&& +{g^2\over 2}\,B_3K_4K_4 + {g^3\over 24} f_7K_4+
     {2\pi G\over g} X_7 H_4.
 \eea
We stress that all terms that involve $B_3$ or $b_2$ in this
formula are fixed by their equations of motion \eref{d} and
\eref{b}; in particular the coefficient of the second term, which
is the unique one involving $b_2$, is fixed  by the
PST--symmetries. There are two terms in $L_{11}$ which are
independent of $B_3$ and $b_2$ and which are not fixed by the
equations of motion, but by the invariance requirements
\eref{1}--\eref{3}: $2\pi G X_7 K_4$ $(a)$, and ${g^3\over 24}
f_7K_4$ $(b)$. The term $(a)$ is related with the contribution
$X_8$ at the r.h.s. of \eref{b}: to get this contribution it would
have been sufficient to include only the term ${2\pi G\over g} X_7
dB_3$ in $L_{11}$ which would have led  to no $SO(1,10)$--anomaly
in $S_{wz}$, since $\int X_7dB_3=\int X_8B_3$ is
$SO(1,10)$--invariant; but the point is that the term ${2\pi
G\over g}X_7dB_3$ alone is not invariant under \eref{1}--\eref{3}
and so one has to add the term $(a)$ (to obtain ${2\pi G\over g}
X_7 H_4$), which introduces in turn an $SO(1,10)$--anomaly.

For the same reason one has to add the term $(b)$; without this
term the first four terms in $L_{11}$ would not be invariant under
\eref{1}--\eref{3}. A straightforward calculation shows, indeed, that
$L_{11}$ as given above is invariant under the transformations
\eref{1}--\eref{3}, as well as under the standard gauge
transformations $\delta B_3=d\Lambda_2$, $\delta
b_2=d\lambda_1-\Lambda_2^{(0)}$, up to a closed form. 

A formal
device to make all these invariances of $S_{wz}$ {\it manifest}
consists in computing the differential of $L_{11}$. Using the formulae 
of the preceding section one obtains  
 \beq
 \label{L12}
L_{12}=dL_{11}={1\over 6}\,H_4H_4H_4+{g\over 2}\left(h_3dB_3^{(0)}
\right)\circ J_5
      +{g^3\over 24}\,P_7J_5+{2\pi G\over g}\left(X_8H_4+gX_7J_5\right).
 \eeq
To give meaning to this formula one has to go to twelve dimensions;
the closed 5--brane has to be extended to a closed 6--brane in $M_{11}
\times R$
with worldvolume $M_7\supset M_6$, in such a way that the
restriction to $M_6$ of the normal bundle of $M_7$ w.r.t. to
$M_{11} \times R$ coincides with the normal bundle of $M_6$ w.r.t.
$M_{11}$. The form $J_5$ is here then the Poincar\`e--dual of
$M_7$ w.r.t. $M_{11} \times R$; restricted to $M_{11}$ it coincides 
with the eleven--dimensional $J_5$ appearing in $L_{11}$.
     
In $L_{12}$ the potentials appear only through their curvatures or 
through $dB_3^{(0)}$, which are all manifestly invariant under 
\eref{1}--\eref{3}. The Chern--Simons form $P_7$ entering in $L_{12}$ 
is defined in terms of the extended $SO(5)$--connection 
$A^{ab}(\sigma,y)$, but since it appears multiplied by $J_5$ one 
gets back $A^{ab}(\sigma,0)=A^{ab}(\sigma)$ and hence also the term 
$P_7J_5$ is independent of the chosen extension. This means that under
\eref{1}--\eref{3} we have $L_{12}^\prime=L_{12}$, and therefore 
$L_{11}^\prime=L_{11}+dL_{10}$ for some ten--form; this ensures
that $S_{wz}$ is invariant. 

From the twelve--dimensional point of view the term ${g^3\over 24}P_7J_5$ 
is necessary to make $L_{12}$ a closed form, as can be seen using 
\eref{first}--\eref{third}. 

It is now easy to compute the gravitational anomalies carried by the
classical action; the kinetic terms are invariant and in the 
Wess--Zumino action only the last two terms contribute, due to 
$\delta f_7=dP_6$, $\delta X_7=dX_6$, with
$$
\delta \left({1\over G}\,S_{wz}\right)=-{2\pi}\int_{M_6}\left(X_6^{(0)}+
{g^3\over 2\pi G}{1\over 24}\,P_6\right).
$$
This should cancel against the quantum anomaly $\delta\Gamma_q$ in
\eref{grav}. To see that this is indeed the case it suffices to
remember that the $5$--brane tension in $M$--theory is tied to
Newton's constant \cite{X8,tensioni} by $T_5=\left({2\pi\over
G^2}\right)^{1\over 3}$. From \eref{kin} and \eref{action} we see
that in our framework the 5--brane tension amounts to $T_5={g\over
G}$. This means that the magnetic charge of the 5--brane is tied
to Newton's constant by
$$
g^3= 2\pi G,
$$
and the effective action is anomaly free. So anomaly cancellation
confirms
once more that there is only one fundamental scale in $M$--theory.

\section{Coupling to $M2$--branes}

It is now simple to couple our action to a closed $M2$--brane with
charge $e$ and
worldvolume $M_3$. If we indicate the current
associated to the 2--brane, i.e. the  Poincar\'e dual of $M_3$,
with $J_8$ ($dJ_8=0$), it is only eq. \eref{b} that gets modified
to
$$
d*H_4={1\over 2}\,H_4H_4+g\,h_3\circ J_5+{2\pi G\over g}\,X_8+e\,J_8.
$$
When 2--branes and 5--branes are
simultaneously present to write an action we must introduce at
least {\it one} Dirac--brane, see e.g. \cite{LM}.  In the Chern--kernel
approach, which avoids
the Dirac--brane for the 5--brane, we must introduce a
Dirac--3--brane, with worldvolume $M_4$, associated to the
2--brane: $\pa M_4=M_3$. Calling the associated current $C_7$ we
have
$$
J_8=dC_7.
$$
To take the new coupling into account it would be sufficient to 
modify the Wess--Zumino action by the term $e\int_{M_3}B_3=
e\int_{M_{11}}B_3J_8=e\int_{M_{11}}dB_3C_7$; but again, to cope
with \eref{1}--\eref{3}, we have to set
$$
S^{(e,g)}_{wz}\equiv S_{wz}+e\int_{M_{11}}H_4C_7.
$$
Under a change of Dirac--brane $M_4\rightarrow M_4+\pa M_5$, we
have $C_7\rightarrow C_7+dC_6$, where $C_6$ is the Poincar\'e dual
of $M_5$. Under such a change the Wess--Zumino action changes by
$\Delta
S^{(e,g)}_{wz}=e\int_{M_{11}}H_4dC_6=-eg\int_{M_{11}}J_5C_6=
-egN$, where the integer $N$ counts the number of intersections
between $M_5$ and $M_6$. The effective action $\Gamma^{(e,g)}
\equiv {1\over G}\left(S_{kin}+S_{wz}^{(e,g)}\right)+\Gamma_q $
changes accordingly by
$$
\Gamma^{(e,g)}\rightarrow \Gamma^{(e,g)}- {eg \over G}\,N,
$$
which is irrelevant if Dirac's condition \eref{Dirac} holds.

This proves that the Dirac--brane is unobservable and that in
$M$--theory elementary $M2$--branes and elementary $M5$--branes
can consistently coexist, in compatibility with gravitational
anomaly cancellation.

\section{Discussion}

The effective action we constructed incorporates $M2$--branes and
$M5$--branes in a consistent way. It is based on the equations of
motion \eref{d} and \eref{b}, and on the definition of the
potentials $B_3$ and $b_2$ according to \eref{c} and \eref{b3}.
The first step was a proof of the consistency of these equations
of motion using a Chern--kernel which codifies the singularities
of $H_4$ {\it near the 5--brane} in an invariant way. Next we wrote
an action which gives rise to these equations of motion, requiring that
the action does not depend on the structure of the Chern--kernel {\it away 
from the 5--brane}. This action is uniquely determined and cancels
automatically the gravitational anomalies.

In the text we supposed that the system of normal coordinates can
be defined globally. In general one is only guaranteed that it can
be defined in a tubular neighborhood of the 5--brane, see e.g. \cite{BT}.
In this situation one can define a $\widetilde K_4$ in this neighborhood 
as in \eref{k4} -- so there it satisfies $d\widetilde K_4=J_5$ -- and try 
to extend it outside as a closed form. For 5--branes for which such a
$\widetilde K_4$ can be extended to the whole target space
our construction holds true. 
In this case the eight--form $\widetilde K_4\widetilde K_4$ is again
closed and since the target space is supposed to be trivial we have
$\widetilde K_4\widetilde K_4={1\over 4}d\widetilde f_7$, for some 
globally defined seven--form. These ingredients are sufficient to write 
down the corresponding effective action, by replacing in 
\eref{kin} and \eref{L11} $K_4\rightarrow \widetilde K_4$,
$f_7\rightarrow \widetilde f_7$. Notice that in a topologically trivial
target space $J_5$ can always be written as the differential of some
four--form; we ask here more, i.e. that this four--form shares with 
$K_4$ the singular behaviour near $M_6$. 

One may ask which are the equations of motion for the coordinates
$x^\mu(\sigma)$ produced by the classical part of our effective action.
The derivation of these equations might show up some problematic aspects,
due to our use of normal coordinates. Notice, however, that this question
is somewhat academic in that only the total action (classical plus quantum)
is anomaly--free. The question whether there
exists a supersymmetric (or $\kappa$--invariant) version of our
action encounters the same fate: since the classical action carries a
gravitational anomaly, its (possible) supersymmetric extension carries
also a supersymmetry anomaly, the so called ``supersymmetric partner";
this means that also the problem of supersymmetry can be stated only for
the total effective action.

Together with the proofs not reported here in \cite{fullpaper}
we will discuss in particular a duality--symmetric formulation, involving
both the three--form $B_3$ and its dual $B_6$ \cite{bbs}, the coupling of our
action to open membranes ending on 5--branes (which carry gravitational
anomalies on their boundaries, too), and the reduction to ten
dimensions.

\bigskip
\paragraph{Acknowledgements.}
The authors thank I. Bandos, L. Bonora,
M. Cariglia, P. Pasti and D. Sorokin for useful discussions. This work was
supported by the
European Commission RTN programme HPRN--CT2000--00131.



\end{document}